\documentclass[prl, twocolumn, amsfonts, amssymb, amsmath,
               showkeys, nofootinbib, superscriptaddress]{revtex4-2}

\usepackage[english]{babel}
\usepackage[utf8]{inputenc}
\usepackage{mathtools}
\usepackage{physics}
\usepackage{booktabs}
\usepackage{siunitx}
\usepackage{graphicx}
\usepackage{dcolumn}
\usepackage{bm}
\usepackage{xcolor}
\usepackage[pdftex,pdftitle={AGN H0},pdfauthor={Kumar et al.}]{hyperref}
\usepackage{multirow}
\usepackage{hhline}
\usepackage{aas_macros}
\usepackage{orcidlink}
\usepackage[table]{xcolor}
\usepackage{makecell}
\usepackage{subcaption}
\usepackage{verbatim}

\newcolumntype{P}[1]{>{\centering\arraybackslash}p{#1}}

\begin{document}

\preprint{APS/123-QED}

\title{A population of LIGO-Virgo-KAGRA mergers happening inside active galactic nuclei}

\author{Dhruv Kumar\orcidlink{0000-0001-8205-0404}}
\affiliation{Beijing Institute of Mathematical Sciences and Applications,
Beijing 101408, China}
\affiliation{Department of Physics, National Institute of Technology
Agartala, Tripura 799046, India}

\affiliation{Institute for Gravitation and the Cosmos, Department of Physics, Pennsylvania State University, University Park, PA 16802, USA}

\author{Alejandro Torres-Orjuela\orcidlink{0000-0002-5467-3505}}
\email[Corresponding author:]{atorreso@bimsa.cn}
\affiliation{Beijing Institute of Mathematical Sciences and Applications,
Beijing 101408, China}

\begin{abstract}
Multimessenger observations of compact binary mergers in active galactic nuclei (AGN) offer unique probes of black hole formation channels and cosmology. We analyse 18 gravitational-wave events from LVK observing runs O3--O4b paired with 28 candidate AGN counterparts, incorporating supermassive black hole (SMBH)-induced environmental redshift corrections and electromagnetic sky-localisation constraints. Of the 28 candidate pairs, 21 are positive-to-strong-favoured (the most compelling being GW190412\_053044 $\leftrightarrow$ J143041.67$+$355703.8 at $\ln\mathcal{B} = +17.35$), three are inconclusive, and four are negative-to-strong-disfavoured. From the 21 positive-to-strong-favoured binary black hole (BBH) AGN pairs, 13 unique associations can be identified as the preferred ones. The cumulative Bayes factor across all 13 BBH-AGN confirmed associations yields $\ln\mathcal{B}_{\rm comb} \approx +81$, establishing strong collective support for AGN-hosted merger associations. Sky localisation is the dominant driver of model selection; at current detector sensitivity, the environmental redshift corrections are neither decisively confirmed nor ruled out, motivating future high-precision multimessenger follow-up.
\end{abstract}

\maketitle


\textit{Introduction.---}The detection of GW150914~\cite{Abbott2016PRL} 
and the subsequent multimessenger observation of 
GW170817~\cite{Abbott2017PRL, Abbott2017ApJL} established GW astronomy 
as a precision tool for probing strong-field gravity. Across four 
observing runs, the LIGO-Vrigo-KAGRA (LVK)\cite{LIGOScientific:2025snk,KAGRA:2023pio,LIGOScientific:2019lzm,Advanced_LIGO_2015,Acernese_2014} network has catalogued nearly 400 compact binary 
candidates, the majority consistent with binary black hole (BBH) 
mergers~\cite{Abbott2019PRX, Abbott2021PRX, Abbott2023PRX, 
LIGOScientific:2025slb,LIGOScientific:2026sit}. Standard parameter estimation treats these systems as 
isolated, vacuum binaries~\cite{Veitch2015PRD, Ashton2019ApJS, 
kumar2025acceleratingparameterestimationparameterized}; however, 
accumulating evidence suggests that a non-negligible fraction may occur 
inside active galactic nuclei (AGN) accretion disks, where the local gravitational environment 
can leave measurable signatures on the detected 
signal~\cite{McKernan2019ApJL, Rowan_2025, Gr_bner_2020, Torres_Orjuela_2023}.

Confirming that BBH mergers occur in AGN disks would carry broad 
implications. Spectroscopic host redshifts paired with GW luminosity 
distances provide independent standard-siren measurements of the Hubble 
constant, complementing the binary neutron star merger-based 
programme~\cite{Abbott2017ApJL}. The mass spectrum of AGN-channel 
mergers encodes the physics of gas-assisted migration traps and 
hierarchical black hole growth~\cite{McKernan_2012, Bellovary_2016}, 
constraining the contribution of dynamical formation channels to the 
observed BBH population. Moreover, the redshift imprinted by the 
central SMBH probes disk structure and orbital dynamics at scales 
inaccessible to electromagnetic observations 
alone~\cite{Chen_2019, Torres_Orjuela_2023}.

\begin{figure*}[htbp]
\centering
\includegraphics[width=0.9\linewidth]{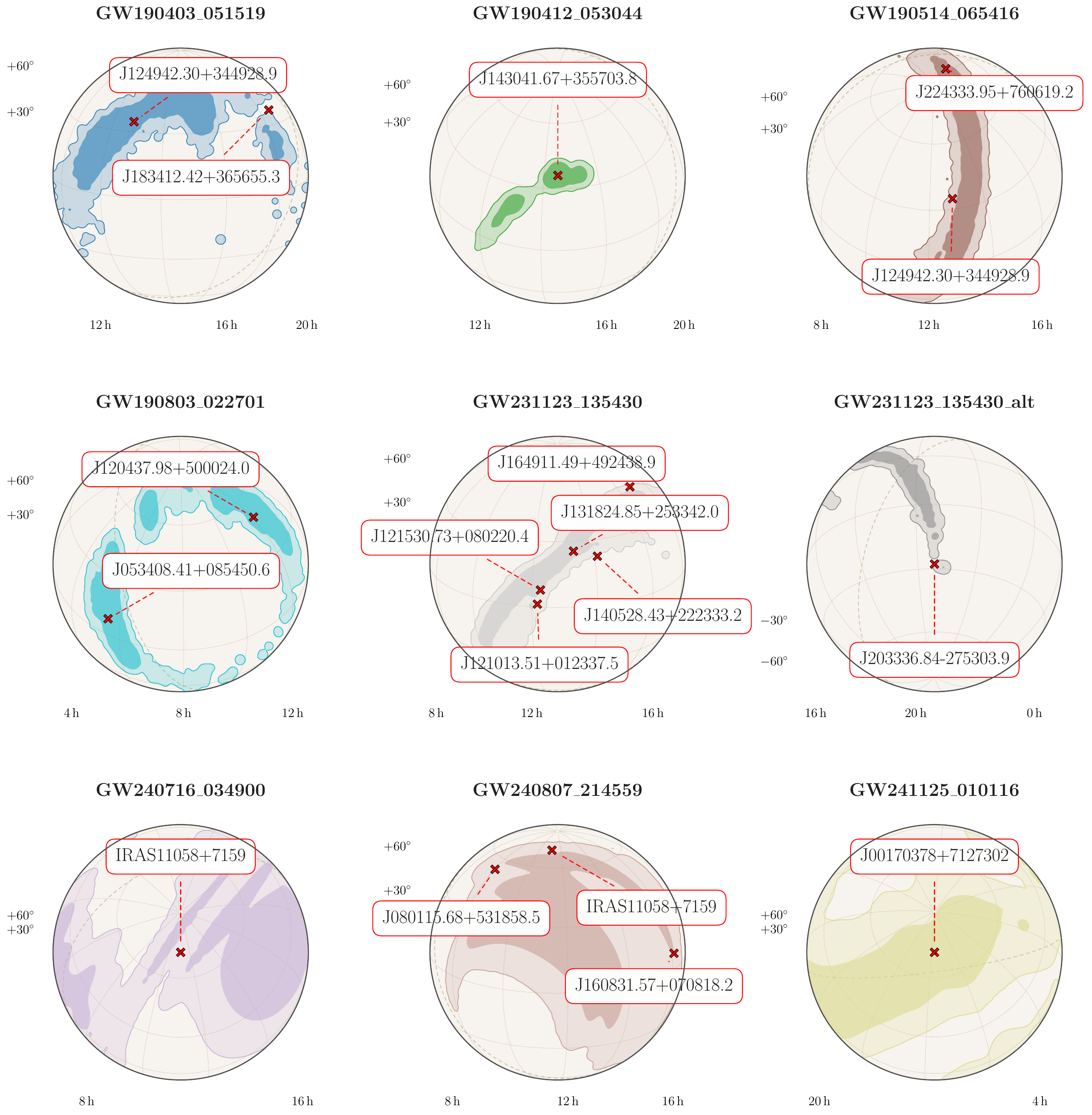}
\caption{GW sky-localisation contours (50\% and 90\% credible regions,
shaded) with candidate AGN positions (red crosses) for a representative
subset of events as given by ~\citet{Graham_2023,cabrera2025searchingelectromagneticemissionagn,bommireddy2026brokerintegratedalgorithmgravitational,gcn38356, Zhang_2026}.}
\label{fig:skymap1}
\end{figure*}

Motivated by these prospects, several groups have searched for candidate 
electromagnetic (EM) counterparts, using spatial correlations -- sky localisation and luminosity distance/cosmological redshift agreements assuming a fixed Cosmology -- and time correlations, as well as the exclusion of other explanations for the EM transients. For O3, Graham et al.~\cite{Graham_2023} reported nine 
coincident AGN flares using the Zwicky Transient Facility at a combined 
$p$-value of $0.0019$, with the GW190521\_030229 counterpart yielding 
$\ln B = 8.6$ in favour of genuine 
association~\cite{morton2023gw190521binaryblackhole}. The sample has 
since grown through O4: Cabrera et 
al.~\cite{cabrera2025searchingelectromagneticemissionagn} identified 
AT\,2023aagj as a candidate for GW230922\_020344, He et 
al.~\cite{he2025searchingelectromagneticcounterpartcandidates} reported 
six candidates for GW231123\_135430, and Bommireddy et 
al.~\cite{bommireddy2026brokerintegratedalgorithmgravitational} flagged 
counterparts for GW240716\_034900, GW240807\_214559, and GW240813\_034548 via automated 
cross-matching. Dedicated follow-up has also targeted 
GW241125\_010116~\cite{gcn38356, Zhang_2026},
S250328ae~\cite{gcn39898, 
zhang2025jointsearchelectromagneticcounterpart} and S251112cm~\cite{vieira2026searchcounterpartsubsolarmass}. Representative sky-localisation contours and AGN positions for a subset
of events are shown in Fig.~\ref{fig:skymap1}.

These searches, however, operate at the photometric level and do not 
directly interrogate what the AGN environment imprints on the GW 
posteriors. In this letter, we construct a marginalised GW likelihood that 
simultaneously accounts for SMBH-induced redshifts and an EM 
sky-localisation constraint, to compute three log Bayes factors 
referenced to the vacuum baseline from four hypothesis classes that 
independently toggle these two ingredients. We apply this framework to 
18 GW events from O3 through O4b paired with 28 candidate AGN 
counterparts compiled from the searches above. The resulting Bayes factors provide the first systematic, GW-side quantification of which proposed counterparts are favoured or disfavoured by the data itself — ranking each candidate's consistency with the GW posteriors rather than constituting an exhaustive search over the full AGN sky.

\textit{Methodology.---}We employ Bayesian inference to extract binary parameters from detector data~\cite{Gregory1992, kumar2025acceleratingparameterestimationparameterized},
using the Bayesian evidence $\mathcal{Z}$ as the model-comparison
statistic~\cite{Gregory1992}, with posterior samples drawn via the
\texttt{nessai} nested-sampling code~\cite{Williams_2021}. All analyses
use the \texttt{IMRPhenomXPHM} waveform approximant~\cite{Pratten_2021,
Roy:2025gzv}, except GW231123\_135430; for this event, we adopt \texttt{IMRPhenomTPHM}~\cite{Estell_s_2022} following Ref.~\cite{Abac_2025_GW231123}, as it better captures the high-mass signature compared to \texttt{IMRPhenomXPHM}. Both waveforms incorporate spin precession and higher-order modes, ensuring a consistent physical treatment across the sample. Binary parameter priors are set per-event from the published LVK configuration~\cite{Veitch2015PRD}, with chirp mass,
mass ratio, and luminosity distance bounds informed by the GWTC posterior
support for each detector-frame signal.

A BBH embedded in an AGN accretion disk and orbiting the central SMBH
accrues two additional redshift contributions beyond the cosmological
term $z_c$: a relativistic Doppler term $z_{\rm rel}$ from the BBH's
orbital speed, and a gravitational term $z_{\rm grav}$ from the SMBH's
potential~\cite{Fabj_2020, Liu_2019, Fang_2019, Chen_2022,
Torres_Orjuela_2023}. Assuming a circular orbit around a non-spinning
SMBH, spin corrections to $z_{\rm grav}$ are negligible at the
separations considered and are omitted~\cite{King_2008, Chen_2022}. The
detected (redshifted) chirp mass is then related to the intrinsic
source-frame mass $M$ by~\cite{Chen_2019}
\begin{equation}
  M_{z_{\rm eff}} = (1+z_c)(1+z_{\rm rel})(1+z_{\rm grav})\,M,
  \label{eq:mass_redshift}
\end{equation}
while aberration of gravitational waves modifies the effective
luminosity distance~\cite{Torres_Orjuela_2023},
\begin{equation}
  D_{\rm eff} = (1+z_{\rm rel})^2\,(1+z_{\rm grav})\,D_L,
  \label{eq:D_eff}
\end{equation}
where $D_L=(1+z_c)\,D_{\rm com}$ is the standard luminosity distance.
Detection of higher spherical-harmonic modes could, in principle, resolve
$z_{\rm rel}$ directly~\cite{Gualtieri_2008, Boyle_2016,
Torres_Orjuela_2021}; however, as detection is dominated by the quadrupolar mode, the degeneracy between mass and redshift remains
unresolved~\cite{Yan_2023}.

To assess the probability of a GW signal originating from an AGN
environment, we construct a joint likelihood combining the GW data with
the EM sky-localisation constraint of the candidate host AGN. Because
the sampler does not natively explore the AGN environmental parameters,
namely the BBH to SMBH separation $r$, the SMBH mass $M_{\rm SMBH}$, and
the angle between the line-of-sight and the binary's center-of-mass
velocity, we marginalise over these at every sampling step,
\begin{align}
  \nonumber \mathcal{L}_{\rm marg}(\Lambda_{\rm GW}) =&
    \int \mathcal{L}_{\rm GW}\!\left(\Lambda_{\rm GW},D_{\rm eff}\right)
    \,\mathcal{L}_{\rm EM}(\alpha,\delta) \\
    &\times\,\pi(\Lambda_{\rm AGN})\;\mathrm{d}\Lambda_{\rm AGN},
  \label{eq:marg_like}
\end{align}
where $\mathcal{L}_{\rm EM}(\alpha,\delta)$ is a 2D Gaussian
sky-localisation likelihood pinned to the exact coordinates of the host
AGN, and $\pi(\Lambda_{\rm AGN})$ is the joint prior on the
environmental parameters. The integral is evaluated by Monte Carlo
marginalisation\cite{Thrane_2019}, drawing $N$ samples directly from
$\pi(\Lambda_{\rm AGN})$ at every sampling step; the marginalised
log-likelihood then reduces to a simple log-mean-exp over likelihood
evaluations,
\begin{align}
  \nonumber \ln\mathcal{L}_{\rm marg} =&
    \ln\!\biggl(\sum_{i=1}^{N}\exp\!\bigl[
      \ln\mathcal{L}_{\rm GW}\!\bigl(D_{{\rm eff},i}\bigr)
      + \ln\mathcal{L}_{\rm EM}(\alpha,\delta)
    \bigr]\biggr) \\
    &- \ln N,
  \label{eq:logsumexp}
\end{align}
which ensures the AGN prior volume is naturally penalised within the
Bayesian evidence $\mathcal{Z}$, favouring localised disk regions where
relativistic corrections are largest.

Because theoretical predictions for the orbital separation at which BBH
mergers occur in AGN disks span a wide range, from migration traps at
$\sim 20$ to $300\,R_s$ where surface density gradients change
sign~\cite{Bellovary_2016} through intermediate disk radii where N-body
simulations confirm rapid binary
formation~\cite{Secunda_2020, McKernan_2012} to the outer disk at
$\gtrsim 10^4\,R_s$~\cite{McKernan_2012}, we adopt a uniform prior on
the BBH to SMBH separation $r$ as the least informative choice, avoiding
bias towards any particular formation model. The $M_{\rm SMBH}$ is drawn
from a truncated log-normal centred on the literature virial mass with
$\sigma = 0.5$~dex, and the angle between the line-of-sight and the
orbital velocity vector is drawn from an isotropic prior,
$p(\theta)\propto\sin\theta$, over $[0,\pi]$, corresponding to a uniform
distribution over the sphere. Because the prior volume on $r$ spans
several orders of magnitude, we partition the separation axis into
overlapping windows to ensure efficient coverage. For runs without the
environmental redshift correction, three far-field windows span
$r/R_s \in [100,\,10{,}000]$; for runs with the correction, six
overlapping windows with widths $\Delta(r/R_s) \in \{2, 4, 8, 16, 32,
64\}$ tile the strong-field range $r/R_s \in [3.1,\,67]$ with 50\%
overlap. The reported log Bayes factor is the median over all windows in
the corresponding ladder. Throughout the analysis, we assume a Planck\,2018
cosmology~\cite{Planck_2020}.

To isolate the individual contributions of the environmental redshift
correction and the EM sky-localisation constraint, we define four
hypothesis classes for each event $\leftrightarrow$ AGN candidate pair:
\begin{itemize}
  \item $\mathcal{H}_{Nz_{\rm eff},\,NEM}$ \textit{(vacuum baseline)}:
        Standard GW-only parameter estimation with $z_c$ drawn as a
        split-normal nuisance parameter from the GW-inferred
        $z_{\rm GW}$ posterior; no redshift correction or EM prior.
  \item $\mathcal{H}_{Nz_{\rm eff},\,YEM}$ \textit{(EM localisation only)}:
        The 2D Gaussian sky prior is centred on the AGN host coordinates
        with $D_{\rm eff} = D_L$ and $z_c$ fixed to $z_{\rm AGN}$; no
        redshift correction.
  \item $\mathcal{H}_{Yz_{\rm eff},\,NEM}$ \textit{(AGN redshift only)}:
        Eq.~\eqref{eq:marg_like} is evaluated with
        $\mathcal{L}_{\rm EM} \equiv 1$ and $z_c$ drawn from the
        $z_{\rm GW}$ posterior; SMBH-induced redshifts modify
        $D_{\rm eff}$ without any EM constraint.
  \item $\mathcal{H}_{Yz_{\rm eff},\,YEM}$ \textit{(full AGN model)}:
        Both the redshift correction and the EM sky prior are applied,
        with $z_c$ fixed to $z_{\rm AGN}$.
\end{itemize}
Pairwise log Bayes factors between these classes decompose the total
evidence into physically distinct contributions:
\begin{itemize}
  \item $\ln\mathcal{B}(\mathrm{NY/NN}) \equiv
        \ln\mathcal{B}^{\mathcal{H}_{Nz_{\rm eff},\,YEM}}_{\mathcal{H}_{Nz_{\rm eff},\,NEM}}$:
        the evidence for EM sky-localisation alone, with no redshift
        correction applied in either hypothesis.
  \item $\ln\mathcal{B}(\mathrm{YY/YN}) \equiv
        \ln\mathcal{B}^{\mathcal{H}_{Yz_{\rm eff},\,YEM}}_{\mathcal{H}_{Yz_{\rm eff},\,NEM}}$:
        the marginal evidence for EM sky-localisation \emph{conditional
        on} the AGN redshift correction already being applied; this
        isolates the sky-prior contribution within the full AGN
        framework.
  \item $\ln\mathcal{B}(\mathrm{YY/NN}) \equiv
        \ln\mathcal{B}^{\mathcal{H}_{Yz_{\rm eff},\,YEM}}_{\mathcal{H}_{Nz_{\rm eff},\,NEM}}$:
        the total evidence for the full AGN model against the vacuum
        baseline, combining the contributions of both the redshift
        correction and the EM prior.
\end{itemize}
All Bayes factors are interpreted following \citet{Kass:1995loi}.


\begin{figure*}[htbp]
\centering
\includegraphics[width=0.75\linewidth]{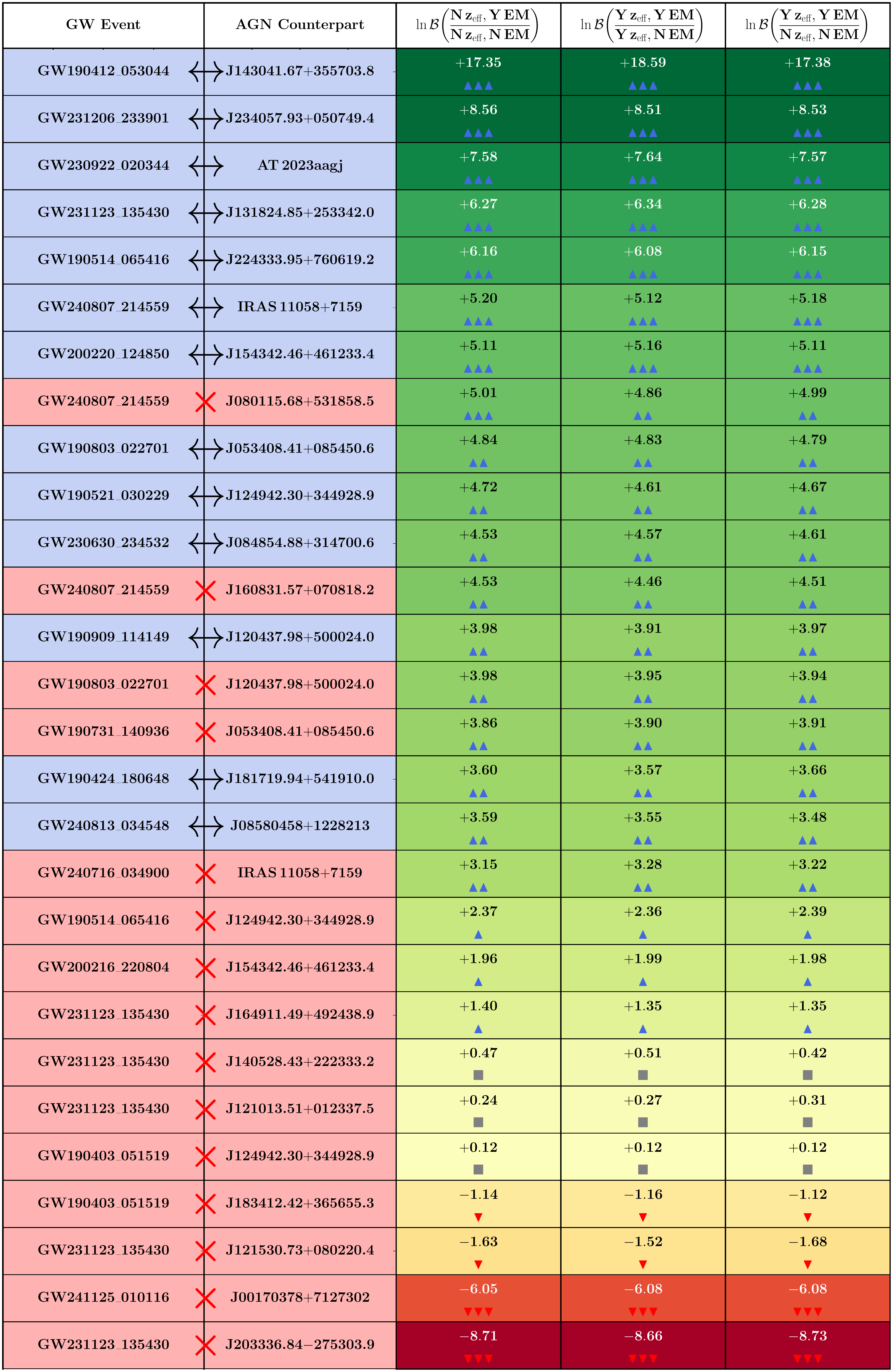}
\caption{Bayes factor matrix for all 28 event $\leftrightarrow$ AGN pairs
across $\ln B(\text{NY/NN})$, $\ln B(\text{YY/YN})$, and
$\ln B(\text{YY/NN})$. Rows are sorted by $\ln B(\text{YY/NN})$ in descending order. Arrows
($\leftrightarrow$) mark the preferred counterpart for each GW event;
crosses ($\textcolor{red}{\times}$) mark non-preferred candidates.
Classification symbols follow \citet{Kass:1995loi}:
$\textcolor{red}{\blacktriangledown\blacktriangledown\blacktriangledown}$
($\ln B < -5$),
$\textcolor{red}{\blacktriangledown\blacktriangledown}$ ($-5$ to $-3$),
$\textcolor{red}{\blacktriangledown}$ ($-3$ to $-1$) indicate preference for the baseline model;
$\textcolor{gray}{\blacksquare}$ ($-1$ to $+1$) is inconclusive;
$\textcolor{blue}{\blacktriangle}$ ($+1$ to $+3$),
$\textcolor{blue}{\blacktriangle\blacktriangle}$ ($+3$ to $+5$), and
$\textcolor{blue}{\blacktriangle\blacktriangle\blacktriangle}$
($\ln B > +5$) indicate preference for the tested model.}
\label{fig:bayes_heatmap}
\end{figure*}

\textit{Results \& Discussion.---}The full Bayes factor matrix for all 28
event $\leftrightarrow$ AGN pairs is shown in Fig.~\ref{fig:bayes_heatmap}. The cumulative sum of $\ln\mathcal{B}(\mathrm{NY/NN})$ across all 28 pairs reaches
$+91.05$ ($+108.58$ from favoured pairs, $-17.53$ from disfavoured),
and $\ln\mathcal{B}(\mathrm{YY/NN})$ tracks it closely at $+90.91$ ($+108.52$
positive, $-17.61$ negative), confirming that the sky-localisation prior
drives the total evidence. The marginal column $\ln\mathcal{B}(\mathrm{YY/YN})$,
measuring the additional contribution of EM localisation within the full
AGN framework, sums to $+92.11$ ($+109.53$ positive, $-17.42$
negative), consistent with NY/NN and confirming that the EM prior
contributes comparably whether or not the redshift correction is applied. Our results show that modeling this set of BBH mergers
as a population inside an AGN disk is strongly preferred over them
being field binaries. However, there is no clear preference between the
model incorporating the redshift effects -- meaning the binary merges close
to the central SMBH -- over the binary merging far from the SMBH.
This does not imply that SMBH-induced redshifts are
absent; rather, the degeneracy between mass and redshift in the dominant
quadrupolar mode prevents the data from isolating a redshift-inflated signal
from one at a slightly different intrinsic mass, while other
corrections are too small at the current detector
sensitivity to shift the evidence appreciably. Spatial coincidence with the host AGN, therefore, remains the only component that presently drives model selection. We also verified that this conclusion is robust to the assumed cosmology: the uncertainty in the GW-inferred luminosity distance is always broad enough to accommodate solutions consistent with the candidate AGN redshift, so switching to an alternative cosmological model does not alter any of the favoured associations.
 
We classify every candidate pair using $\ln\mathcal{B}(\mathrm{YY/NN})$ as
follows: a pair is \emph{favoured} if it is one of the
thirteen events marked in blue in Fig.~\ref{fig:bayes_heatmap} obtained
after resolving all competing candidates, while a candidate is \emph{competitively disfavoured}
if it satisfies $\ln\mathcal{B}>1$ is non-preferred -- individually
consistent with its host AGN, but out-competed by an alternative pairing;
candidates with $-1<\ln\mathcal{B}<1$ are \emph{inconclusive}; any candidate with $\ln\mathcal{B}<-1$ is \emph{disfavoured} and any candidate with $\ln\mathcal{B}<-5$ is \emph{strongly disfavoured}.
 
Of the eighteen GW events analysed, thirteen propose only a single
candidate AGN counterpart. Of these, nine are strongly or very strongly
favoured. GW190412\_053044 $\leftrightarrow$
J143041.67+355703.8 yields $\ln\mathcal{B}(\mathrm{YY/NN}) = +17.38$, the most
significant in the sample, confirming the ZTF flare identification of
\citet{Zhu__2026}. GW231206\_233901 $\leftrightarrow$
J234057.93+050749.4 ($+8.53$) is consistent with
\citet{darc2025longtermopticalfollows231206cc}, and GW230922\_020344
$\leftrightarrow$ AT\,2023aagj ($+7.57$) confirms the counterpart of
\citet{cabrera2025searchingelectromagneticemissionagn}. Six other events are also strongly favoured, yielding log-Bayes factors between 3.48 and 5.11. The remaining four single-candidate events are not retained as the
preferred association. The pair GW241125\_010116 $\leftrightarrow$ J00170378+7127302 is strongly disfavoured at $\ln\mathcal{B}(\mathrm{YY/NN}) = -6.08$ -- driven by the large discrepancy between the GW-inferred and AGN spectroscopic
redshifts ($z_{\rm GW} = 0.744$ versus $z_{\rm AGN} = 0.170$) -- while the other three candidates are competitively disfavoured, as a different event independently claims the same AGN with higher evidence.

For events with multiple candidate counterparts, our framework provides
clear rankings.
GW231123\_135430 has six candidates, where J131824.85+253342.0 is favoured 
($+6.28$) and J164911.49+492438.9 ($+1.35$) is competitively disfavoured, while two AGN candidates are inconclusive and two are strongly disfavoured. In particular, the AGN candidate J203336.84$-$275303.9 is strongly disfavoured ($-8.73$), as the AGN falls entirely outside the GW (90\,\%) localisation contour and because the large
discrepancy between the GW-inferred and AGN spectroscopic redshifts
($z_{\rm GW} = 0.465$ versus $z_{\rm AGN} = 0.149$) prevents the
luminosity distances from being reconciled. That the same event
simultaneously yields $+6.28$ for one candidate and $-8.73$ for another
demonstrates that the framework reliably discriminates between competing
counterparts. GW190514\_065416 has two candidates:
J224333.95+760619.2 ($+6.15$) being favoured over
J124942.30+344928.9 ($+2.39$).
independently claimed more strongly by GW190521\_030229).
GW190803\_022701 also has two candidates, where
J053408.41+085450.6 ($+4.79$) is favoured over J120437.98+500024.0
($+3.94$).
more strongly by GW190909\_114149). GW240807\_214559 has three candidates, all being individually substantial:
IRAS\,11058+7159 ($+5.18$) is favoured, leading over J080115.68+531858.5 ($+4.99$)
and J160831.57+070818.2 ($+4.51$). For
GW190403\_051519, neither candidate is favoured:
J124942.30+344928.9 ($+0.12$) is inconclusive, and
J183412.42+365655.3 ($-1.12$) is strongly disfavoured.
 
A complementary test arises when the same AGN is independently claimed
by two different GW events: here the framework must correctly assign one
host to one of two candidate mergers, rather than one merger to one of
several candidate hosts. Five such AGN occur in our sample, and in every
case the higher-evidence pairing is favoured while the lower one is
competitively disfavoured: J124942.30+344928.9 is awarded to
GW190521\_030229 ($+4.67$) over GW190514\_065416 ($+2.39$),
J053408.41+085450.6 to GW190803\_022701 ($+4.79$) over
GW190731\_140936 ($+3.91$), J120437.98+500024.0 to
GW190909\_114149 ($+3.97$) over GW190803\_022701 ($+3.94$),
J154342.46+461233.4 to GW200220\_124850 ($+5.11$) over
GW200216\_220804 ($+1.98$), and IRAS\,11058+7159 to
GW240807\_214559 ($+5.18$) over GW240716\_034900 ($+3.22$). Notably,
GW190803\_022701 plays both roles at once: it is the favoured host
assignment for J053408.41+085450.6, competitively disfavouring
GW190731\_140936, while its own second candidate,
J120437.98+500024.0, is itself competitively disfavoured against
GW190909\_114149. Together with the GW-side multiplicity discussed
above, this demonstrates that our pipeline performs reliable coincidence
classification in both directions -- one merger with several candidate
hosts, and one host with several candidate mergers -- converging on a
single, internally consistent set of thirteen favoured associations.

Four of the thirteen favoured pairs show marginal near-SMBH tendencies, with $\ln\mathcal{B}(\mathrm{YY/NN})$
exceeding $\ln\mathcal{B}(\mathrm{NY/NN})$ and $\ln\mathcal{B}(\mathrm{YY/YN})$ comparable to
$\ln\mathcal{B}(\mathrm{NY/NN})$, indicating a slight preference for a merger occurring close to the central SMBH, consistent with AGN migration trap predictions~\cite{Bellovary_2016, McKernan_2012}.
 
These measurements remain individually sub-threshold because the environmental
corrections at the sampled separations are too small to produce
distinguishable posterior shifts at the current SNR. However, several
complementary observables could resolve this in future analyses:
detection of higher spherical-harmonic modes, which encode the redshift
independently of the chirp mass~\cite{Gualtieri_2008, Boyle_2016,
Torres_Orjuela_2021}; measurement of the centre-of-mass acceleration
imprinted on the GW phase by the SMBH orbital
motion~\cite{inayoshi_tamanini_2017, tamanini_klein_2019,
Torres_Orjuela_2023, vijaykumar_tiwari_2023, lazarow_leslie_2024,
yang_han_2025, zhao_yan_2026}; and time-delay signatures between the GW
signal and an EM counterpart~\cite{Haiman_2017, Chen_2019}. Improved sky
localisation in future runs will further sharpen the test: for AGN that
remain within the tighter GW error box, the EM prior will saturate and
the environmental redshift correction will become the factor that
distinguishes a chance positional coincidence from a genuine
disk-embedded merger.
 
We have analysed 18 gravitational-wave events from LVK observing runs
O3--O4b paired with 28 candidate AGN counterparts, computing three
pairwise log Bayes factors that independently quantify the contributions
of EM sky-localisation and SMBH-induced environmental redshifts against
a vacuum baseline. The key findings are:
\emph{(i)~Evidence}: 21 of 28 candidate pairs
have $\ln\mathcal{B}(\mathrm{YY/NN})>1$, comprising the thirteen favoured
associations together with eight competitively disfavoured candidates
that remain individually consistent with their host AGN but are
out-competed by an alternative pairing, with the most compelling case
GW190412\_053044 $\leftrightarrow$ J143041.67$+$355703.8 reaching
$\ln\mathcal{B}=+17.38$. The cumulative evidence across the thirteen favoured events yields
$\ln\mathcal{B}_{\rm comb}(\mathrm{YY/NN})\approx+81$, establishing
strong collective support for AGN-hosted merger associations. \emph{(ii)~Discrimination}: for the four GW
events with multiple candidate counterparts
and symmetrically for the five AGN independently claimed by two different
GW events, the framework identifies a single preferred association
in every case, demonstrating that
coincidence classification resolves counterpart ambiguity in both
directions without
additional EM follow-up. \emph{(iii)~Rejection}: four
candidates are disfavoured or strongly disfavoured ($\ln\mathcal{B}<-1$) 
driven by irreconcilable redshift
mismatches and in one case sky-position incompatibility.
\emph{(iv)~Sky localisation vs.\ redshift}: $\ln\mathcal{B}(\mathrm{NY/NN})$
and $\ln\mathcal{B}(\mathrm{YY/NN})$ are nearly identical in cumulative
sum ($+81.49$ and $+81.38$ respectively), while
$\ln\mathcal{B}(\mathrm{YY/YN})$ is consistent with both ($+82.48$),
confirming that sky localisation is the dominant driver of model
selection and that SMBH-induced environmental redshifts are neither
confirmed nor ruled out at current detector sensitivity.
 
The reliable discrimination between competing counterparts, on both the
GW side and the AGN side, and the
conclusive rejection of positionally incompatible candidates establish
this Bayesian coincidence-classification framework as a robust tool for ranking and validating pre-identified AGN counterpart candidates across present and future LVK observing runs.

\textit{Acknowledgment.---}We thank Anna Liu, Bangalore S. Sathyaprakash, Bernard F. Schutz,
and Juan Calder{\'o}n Bustillo for helpful comments and questions.
This work was supported by the National Science Foundation of China
(No.\ W2533010).  This work used data from the LIGO–Virgo–KAGRA
Collaboration provided by the Gravitational Wave Open Science Center
(\url{gwosc.org}).  LIGO is funded by the NSF; Virgo by CNRS, INFN,
and Nikhef through EGO; KAGRA by MEXT and JSPS (Japan), NRF and MSIT
(Korea), and NSTC (Taiwan).  D.K.\ would also like to acknowledge the LIGO Laboratory computing resources supported by NSF grants:PHY-0757058 and PHY-0823459, and the Gwave cluster maintained by the Institute for Computational and Data Sciences at Penn State University, supported by NSF grants: OAC-2346596, OAC-2201445, OAC- 2103662, OAC-2018299, and PHY-2110594. This material is based upon work supported by NSF’s LIGO Laboratory which is a major facility fully funded by the National Science Foundation.


\bibliography{bibliography}

@article{Gregory1992,
  author    = {P. C. Gregory and others},
  title     = {A new method for the detection of a periodic signal of unknown shape and period},
  journal   = {The Astrophysical Journal},
  year      = {1992},
  volume    = {398},
  pages     = {146},
  doi       = {10.1086/171844},
  url       = {https://doi.org/10.1086/171844}
}

@misc{bommireddy2026brokerintegratedalgorithmgravitational,
  title={A Broker Integrated Algorithm for Gravitational Wave - Electromagnetic Counterpart Searches in O4a and O4b Runs}, 
  author={Hemanth Bommireddy and others},
  year={2026},
  eprint={2603.04342},
  archivePrefix={arXiv},
  primaryClass={astro-ph.HE},
  url={https://arxiv.org/abs/2603.04342}, 
}

@misc{vieira2026searchcounterpartsubsolarmass,
  title={Search For a Counterpart to the Subsolar Mass Gravitational Wave Candidate S251112cm}, 
  author={Nicholas Vieira and others},
  year={2026},
  eprint={2603.17009},
  archivePrefix={arXiv},
  primaryClass={astro-ph.HE},
  url={https://arxiv.org/abs/2603.17009}, 
}

@misc{darc2025longtermopticalfollows231206cc,
  title={Long-Term Optical Follow Up of S231206cc: Multi-Model Constraints on BBH Merger Emission in AGN Disks}, 
  author={P. Darc and others},
  year={2025},
  eprint={2506.02224},
  archivePrefix={arXiv},
  primaryClass={astro-ph.HE},
  url={https://arxiv.org/abs/2506.02224}, 
}

@misc{zhang2025jointsearchelectromagneticcounterpart,
  title={A Joint Search for the Electromagnetic Counterpart to the Gravitational-Wave Binary Black-Hole Merger Candidate S250328ae with the Dark Energy Camera and the Prime Focus Spectrograph}, 
  author={Haibin Zhang and others},
  year={2025},
  eprint={2508.00291},
  archivePrefix={arXiv},
  primaryClass={astro-ph.HE},
  url={https://arxiv.org/abs/2508.00291}, 
}

@article{Zhang_2026,
  title={LVK S241125n: Massive Binary Black Hole Merger Produces Gamma Ray Burst in Active Galactic Nucleus Disk},
  volume={998},
  ISSN={1538-4357},
  url={http://dx.doi.org/10.3847/1538-4357/ae3319},
  DOI={10.3847/1538-4357/ae3319},
  number={1},
  journal={The Astrophysical Journal},
  publisher={American Astronomical Society},
  author={Zhang, Shu-Rui and others},
  year={2026},
  month=feb, pages={171} }

@misc{he2025searchingelectromagneticcounterpartcandidates,
  title={Searching for Electromagnetic Counterpart Candidates to GW231123}, 
  author={Lei He and others},
  year={2025},
  eprint={2511.05144},
  archivePrefix={arXiv},
  primaryClass={astro-ph.HE},
  url={https://arxiv.org/abs/2511.05144}, 
}

@misc{cabrera2025searchingelectromagneticemissionagn,
  title={Searching for electromagnetic emission in an AGN from the gravitational wave binary black hole merger candidate S230922g}, 
  author={Tomás Cabrera and others},
  year={2025},
  eprint={2407.10698},
  archivePrefix={arXiv},
  primaryClass={astro-ph.HE},
  url={https://arxiv.org/abs/2407.10698}, 
}

@article{Zhu__2026,
  title={Constraining the Fraction of LIGO/Virgo/KAGRA Binary Black Hole Merger Events Associated with Active Galactic Nucleus Flares},
  volume={1000},
  ISSN={1538-4357},
  url={http://dx.doi.org/10.3847/1538-4357/ae47ea},
  DOI={10.3847/1538-4357/ae47ea},
  number={1},
  journal={The Astrophysical Journal},
  publisher={American Astronomical Society},
  author={Zhu, Liang-Gui and others},
  year={2026},
  month=mar, pages={115} }

@misc{kumar2025acceleratingparameterestimationparameterized,
  title={Accelerating parameter estimation for parameterized tests of general relativity with gravitational-wave observations}, 
  author={Dhruv Kumar and others},
  year={2025},
  eprint={2511.16879},
  archivePrefix={arXiv},
  primaryClass={gr-qc},
  url={https://arxiv.org/abs/2511.16879}, 
}

@misc{gcn38356, 
  author = {Piotrzkowski, B. and others},
  title = {{LIGO/Virgo/KAGRA S241125n: Update on Coincidence False Alarm Method}},
  howpublished = {GCN Circular 38356},
  year = {2024},
  month = nov,
  url = {https://gcn.nasa.gov/circulars/38356},
}

@misc{gcn39898,
  author = {Ghosh, S. and others},
  title = {{LIGO/Virgo/KAGRA S250328ae: Identification of a GW Compact Binary Merger Candidate}},
  howpublished = {GCN Circular 39898},
  year = {2025},
  month = mar,
  url = {https://gcn.nasa.gov/circulars/39898},
}

@article{Acernese_2014,
   title={Advanced Virgo: a second-generation interferometric gravitational wave detector},
   volume={32},
   ISSN={1361-6382},
   url={http://dx.doi.org/10.1088/0264-9381/32/2/024001},
   DOI={10.1088/0264-9381/32/2/024001},
   number={2},
   journal={Classical and Quantum Gravity},
   publisher={IOP Publishing},
   author={Acernese, F and others},
   year={2014},
   month=Dec, pages={024001} }

@article{Advanced_LIGO_2015,
   title={Advanced LIGO},
   volume={32},
   ISSN={1361-6382},
   url={http://dx.doi.org/10.1088/0264-9381/32/7/074001},
   DOI={10.1088/0264-9381/32/7/074001},
   number={7},
   journal={Classical and Quantum Gravity},
   publisher={IOP Publishing},
   author={ and Aasi, J and others},
   year={2015},
   month=Mar, pages={074001} }

@article{Bellovary_2016,
  title={MIGRATION TRAPS IN DISKS AROUND SUPERMASSIVE BLACK HOLES},
  volume={819},
  ISSN={2041-8213},
  url={http://dx.doi.org/10.3847/2041-8205/819/2/L17},
  DOI={10.3847/2041-8205/819/2/l17},
  number={2},
  journal={The Astrophysical Journal Letters},
  publisher={American Astronomical Society},
  author={Bellovary, Jillian M. and others},
  year={2016},
  month=Mar, pages={L17} }

@article{McKernan_2012,
  title={Intermediate mass black holes in AGN discs - I. Production and growth: IMBH in AGN discs - I},
  volume={425},
  ISSN={0035-8711},
  url={http://dx.doi.org/10.1111/j.1365-2966.2012.21486.x},
  DOI={10.1111/j.1365-2966.2012.21486.x},
  number={1},
  journal={Monthly Notices of the Royal Astronomical Society},
  publisher={Oxford University Press (OUP)},
  author={McKernan, B. and others},
  year={2012},
  month=July, pages={460–469} }

@article{Fabj_2020,
  title={Aligning nuclear cluster orbits with an active galactic nucleus accretion disc},
  volume={499},
  ISSN={1365-2966},
  url={http://dx.doi.org/10.1093/mnras/staa3004},
  DOI={10.1093/mnras/staa3004},
  number={2},
  journal={Monthly Notices of the Royal Astronomical Society},
  publisher={Oxford University Press (OUP)},
  author={Fabj, Gaia and others},
  year={2020},
  month=oct, pages={2608–2616} }

@article{King_2008,
  title={The evolution of black hole mass and spin in active galactic nuclei},
  volume={385},
  ISSN={1365-2966},
  url={http://dx.doi.org/10.1111/j.1365-2966.2008.12943.x},
  DOI={10.1111/j.1365-2966.2008.12943.x},
  number={3},
  journal={Monthly Notices of the Royal Astronomical Society},
  publisher={Oxford University Press (OUP)},
  author={King, A. R. and others},
  year={2008},
  month=apr, pages={1621–1627} }

@article{Fang_2019,
  title={Secular evolution of compact binaries revolving around a spinning massive black hole},
  volume={99},
  ISSN={2470-0029},
  url={http://dx.doi.org/10.1103/PhysRevD.99.103005},
  DOI={10.1103/physrevd.99.103005},
  number={10},
  journal={Physical Review D},
  publisher={American Physical Society (APS)},
  author={Fang, Yun and others},
  year={2019},
  month=may }

@article{Liu_2019,
  title={Binary Mergers near a Supermassive Black Hole: Relativistic Effects in Triples},
  volume={883},
  ISSN={2041-8213},
  url={http://dx.doi.org/10.3847/2041-8213/ab40c0},
  DOI={10.3847/2041-8213/ab40c0},
  number={1},
  journal={The Astrophysical Journal Letters},
  publisher={American Astronomical Society},
  author={Liu, Bin and others},
  year={2019},
  month=sep, pages={L7} }

@article{Chen_2022,
  title={Binaries wandering around supermassive black holes due to gravitoelectromagnetism},
  volume={106},
  ISSN={2470-0029},
  url={http://dx.doi.org/10.1103/PhysRevD.106.103040},
  DOI={10.1103/physrevd.106.103040},
  number={10},
  journal={Physical Review D},
  publisher={American Physical Society (APS)},
  author={Chen, Xian and others},
  year={2022},
  month=nov }

@article{Chen_2019,
  title={Mass–redshift degeneracy for the gravitational-wave sources in the vicinity of supermassive black holes},
  volume={485},
  ISSN={1745-3933},
  url={http://dx.doi.org/10.1093/mnrasl/slz046},
  DOI={10.1093/mnrasl/slz046},
  number={1},
  journal={Monthly Notices of the Royal Astronomical Society: Letters},
  publisher={Oxford University Press (OUP)},
  author={Chen, Xian and others},
  year={2019},
  month=apr, pages={L141–L145} }

@article{Torres_Orjuela_2023,
  title={Moving gravitational wave sources at cosmological distances: Impact on the measurement of the Hubble constant},
  volume={107},
  ISSN={2470-0029},
  url={http://dx.doi.org/10.1103/PhysRevD.107.043027},
  DOI={10.1103/physrevd.107.043027},
  number={4},
  journal={Physical Review D},
  publisher={American Physical Society (APS)},
  author={Torres-Orjuela, Alejandro and others},
  year={2023},
  month=feb }

@article{Gualtieri_2008,
  title={Transformation of the multipolar components of gravitational radiation under rotations and boosts},
  volume={78},
  ISSN={1550-2368},
  url={http://dx.doi.org/10.1103/PhysRevD.78.044024},
  DOI={10.1103/physrevd.78.044024},
  number={4},
  journal={Physical Review D},
  publisher={American Physical Society (APS)},
  author={Gualtieri, Leonardo and others},
  year={2008},
  month=aug }

@article{Boyle_2016,
  title={Transformations of asymptotic gravitational-wave data},
  volume={93},
  ISSN={2470-0029},
  url={http://dx.doi.org/10.1103/PhysRevD.93.084031},
  DOI={10.1103/physrevd.93.084031},
  number={8},
  journal={Physical Review D},
  publisher={American Physical Society (APS)},
  author={Boyle, Michael},
  year={2016},
  month=apr }

@article{Torres_Orjuela_2021,
  title={Excitation of gravitational wave modes by a center-of-mass velocity of the source},
  volume={104},
  ISSN={2470-0029},
  url={http://dx.doi.org/10.1103/PhysRevD.104.123025},
  DOI={10.1103/physrevd.104.123025},
  number={12},
  journal={Physical Review D},
  publisher={American Physical Society (APS)},
  author={Torres-Orjuela, Alejandro and others},
  year={2021},
  month=dec }

@article{Yan_2023,
  title={Calculating the gravitational waves emitted from high-speed sources},
  volume={107},
  ISSN={2470-0029},
  url={http://dx.doi.org/10.1103/PhysRevD.107.103044},
  DOI={10.1103/physrevd.107.103044},
  number={10},
  journal={Physical Review D},
  publisher={American Physical Society (APS)},
  author={Yan, Han and others},
  year={2023},
  month=may }

@misc{morton2023gw190521binaryblackhole,
  title={GW190521: a binary black hole merger inside an active galactic nucleus?}, 
  author={Sophia Morton and others},
  year={2023},
  eprint={2310.16025},
  archivePrefix={arXiv},
  primaryClass={gr-qc},
  url={https://arxiv.org/abs/2310.16025}, 
}

@article{Graham_2023,
  title={A Light in the Dark: Searching for Electromagnetic Counterparts to Black Hole–Black Hole Mergers in LIGO/Virgo O3 with the Zwicky Transient Facility},
  volume={942},
  ISSN={1538-4357},
  url={http://dx.doi.org/10.3847/1538-4357/aca480},
  DOI={10.3847/1538-4357/aca480},
  number={2},
  journal={The Astrophysical Journal},
  publisher={American Astronomical Society},
  author={Graham, Matthew J. and others},
  year={2023},
  month=jan, pages={99} }

@article{Abac_2025_GW231123,
   title={GW231123: A Binary Black Hole Merger with Total Mass 190–265
                    <i>M</i>
                    <sub>⊙</sub>},
   volume={993},
   ISSN={2041-8213},
   url={http://dx.doi.org/10.3847/2041-8213/ae0c9c},
   DOI={10.3847/2041-8213/ae0c9c},
   number={1},
   journal={The Astrophysical Journal Letters},
   publisher={American Astronomical Society},
   author={Abac, A. G. and others},
   year={2025},
   month=Oct, pages={L25} }

@article{Estell_s_2022,
   title={New twists in compact binary waveform modeling: A fast time-domain model for precession},
   volume={105},
   ISSN={2470-0029},
   url={http://dx.doi.org/10.1103/PhysRevD.105.084040},
   DOI={10.1103/physrevd.105.084040},
   number={8},
   journal={Physical Review D},
   publisher={American Physical Society (APS)},
   author={Estellés, Héctor and others},
   year={2022},
   month=Apr }

@article{Thrane_2019,
   title={An introduction to Bayesian inference in gravitational-wave astronomy: Parameter estimation, model selection, and hierarchical models},
   volume={36},
   ISSN={1448-6083},
   url={http://dx.doi.org/10.1017/pasa.2019.2},
   DOI={10.1017/pasa.2019.2},
   journal={Publications of the Astronomical Society of Australia},
   publisher={Cambridge University Press (CUP)},
   author={Thrane, Eric and Talbot, Colm},
   year={2019} }

@misc{LIGOScientific:2026sit,
      title={GWTC-5.0: An Introduction to Version 5.0 of the Gravitational-Wave Transient Catalog}, 
      author={LIGO Scientific Collaboration and the Virgo Collaboration and the KAGRA Collaboration and A. G. Abac and others},
      year={2026},
      eprint={2605.27223},
      archivePrefix={arXiv},
      primaryClass={gr-qc},
      url={https://arxiv.org/abs/2605.27223}, 
}

@article{Pratten_2021,
  title={Computationally efficient models for the dominant and subdominant harmonic modes of precessing binary black holes},
  volume={103},
  ISSN={2470-0029},
  url={http://dx.doi.org/10.1103/PhysRevD.103.104056},
  DOI={10.1103/physrevd.103.104056},
  number={10},
  journal={Physical Review D},
  publisher={American Physical Society (APS)},
  author={Pratten, Geraint and others},
  year={2021},
  month=may }

@article{Roy:2025gzv,
    author = "Roy, Soumen and Haney, Maria and Pratten, Geraint and T. H. Pang, Peter and Van Den Broeck, Chris",
    title = "{Improved parametrized test of general relativity using the IMRPhenomX waveform family: Including higher harmonics and precession}",
    eprint = "2504.21147",
    archivePrefix = "arXiv",
    primaryClass = "gr-qc",
    reportNumber = "LIGO DCC P2500034",
    doi = "10.1103/855k-sys5",
    journal = "Phys. Rev. D",
    volume = "113",
    number = "2",
    pages = "024016",
    year = "2026"
}

@article{Williams_2021,
  title={Nested sampling with normalizing flows for gravitational-wave inference},
  volume={103},
  ISSN={2470-0029},
  url={http://dx.doi.org/10.1103/PhysRevD.103.103006},
  DOI={10.1103/physrevd.103.103006},
  number={10},
  journal={Physical Review D},
  publisher={American Physical Society (APS)},
  author={Williams, Michael J. and others},
  year={2021},
  month=May }

@article{Abbott2016PRL,
  author = "Abbott, B. P. and others",
  collaboration = "LIGO Scientific, Virgo",
  title = "{Observation of Gravitational Waves from a Binary Black Hole Merger}",
  eprint = "1602.03837",
  archivePrefix = "arXiv",
  primaryClass = "gr-qc",
  doi = "10.1103/PhysRevLett.116.061102",
  journal = "Phys. Rev. Lett.",
  volume = "116",
  number = "6",
  pages = "061102",
  year = "2016"
}

@article{Abbott2017PRL,
  author = "Abbott, B. P. and others",
  collaboration = "LIGO Scientific, Virgo",
  title = "{GW170817: Observation of Gravitational Waves from a Binary Neutron Star Inspiral}",
  eprint = "1710.05832",
  archivePrefix = "arXiv",
  primaryClass = "gr-qc",
  doi = "10.1103/PhysRevLett.119.161101",
  journal = "Phys. Rev. Lett.",
  volume = "119",
  number = "16",
  pages = "161101",
  year = "2017"
}

@article{Abbott2017ApJL,
  author = "Abbott, B. P. and others",
  collaboration = "LIGO Scientific, Virgo, Fermi-GBM, INTEGRAL",
  title = "{Multi-messenger Observations of a Binary Neutron Star Merger}",
  eprint = "1710.05833",
  archivePrefix = "arXiv",
  primaryClass = "astro-ph.HE",
  doi = "10.3847/2041-8213/aa91c9",
  journal = "Astrophys. J. Lett.",
  volume = "848",
  number = "2",
  pages = "L12",
  year = "2017"
}

@article{Abbott2019PRX,
  author = "Abbott, B. P. and others",
  collaboration = "LIGO Scientific, Virgo",
  title = "{GWTC-1: A Gravitational-Wave Transient Catalog of Compact Binary Mergers Observed by LIGO and Virgo during the First and Second Observing Runs}",
  eprint = "1811.12907",
  archivePrefix = "arXiv",
  primaryClass = "astro-ph.HE",
  doi = "10.1103/PhysRevX.9.031040",
  journal = "Phys. Rev. X",
  volume = "9",
  number = "3",
  pages = "031040",
  year = "2019"
}

@article{Abbott2021PRX,
  author = "Abbott, R. and others",
  collaboration = "LIGO Scientific, Virgo",
  title = "{GWTC-2: Compact Binary Coalescences Observed by LIGO and Virgo during the First Half of the Third Observing Run}",
  eprint = "2010.14527",
  archivePrefix = "arXiv",
  primaryClass = "gr-qc",
  doi = "10.1103/PhysRevX.11.021053",
  journal = "Phys. Rev. X",
  volume = "11",
  number = "2",
  pages = "021053",
  year = "2021"
}

@article{Abbott2023PRX,
  author = "Abbott, R. and others",
  collaboration = "LIGO Scientific, VIRGO, KAGRA",
  title = "{GWTC-3: Compact Binary Coalescences Observed by LIGO and Virgo during the Second Part of the Third Observing Run}",
  eprint = "2111.03606",
  archivePrefix = "arXiv",
  primaryClass = "gr-qc",
  doi = "10.1103/PhysRevX.13.041039",
  journal = "Phys. Rev. X",
  volume = "13",
  number = "4",
  pages = "041039",
  year = "2023"
}

@article{LIGOScientific:2025slb,
    author = "Abac, A. G. and others",
    collaboration = "LIGO Scientific, VIRGO, KAGRA",
    title = "{GWTC-4.0: Updating the Gravitational-Wave Transient Catalog with Observations from the First Part of the Fourth LIGO-Virgo-KAGRA Observing Run}",
    eprint = "2508.18082",
    archivePrefix = "arXiv",
    primaryClass = "gr-qc",
    reportNumber = "LIGO-P2400386",
    doi = "10.3847/2041-8213/ae2c74",
    journal = "Astrophys. J. Lett.",
    volume = "1004",
    number = "2",
    pages = "L22",
    year = "2026"
}

@article{Veitch2015PRD,
  author = "Veitch, J. and others",
  title = "{Parameter estimation for compact binaries with ground-based gravitational-wave observations using the LALInference software library}",
  eprint = "1409.7215",
  archivePrefix = "arXiv",
  primaryClass = "gr-qc",
  doi = "10.1103/PhysRevD.91.042003",
  journal = "Phys. Rev. D",
  volume = "91",
  number = "4",
  pages = "042003",
  year = "2015"
}

@article{Ashton2019ApJS,
  author = "Ashton, G. and others",
  title = "{BILBY: A User-friendly Bayesian Inference Library for Gravitational-wave Astronomy}",
  eprint = "1811.02042",
  archivePrefix = "arXiv",
  primaryClass = "astro-ph.IM",
  doi = "10.3847/1538-4365/ab06fc",
  journal = "Astrophys. J. Suppl. Ser.",
  volume = "241",
  number = "2",
  pages = "27",
  year = "2019"
}

@article{McKernan2019ApJL,
  author = "McKernan, B. and others",
  title = "{Ram-pressure Stripping of a Kicked Hill Sphere: Prompt Electromagnetic Emission from the Merger of Stellar Mass Black Holes in an AGN Accretion Disk}",
  eprint = "1907.03746",
  archivePrefix = "arXiv",
  primaryClass = "astro-ph.HE",
  doi = "10.3847/2041-8213/ab4886",
  journal = "Astrophys. J. Lett.",
  volume = "884",
  number = "2",
  pages = "L50",
  year = "2019"
}

@article{Rowan_2025,
  title={Black hole merger rates in AGN: contribution from gas-captured binaries},
  volume={544},
  ISSN={1365-2966},
  url={http://dx.doi.org/10.1093/mnras/staf1896},
  DOI={10.1093/mnras/staf1896},
  number={4},
  journal={Monthly Notices of the Royal Astronomical Society},
  publisher={Oxford University Press (OUP)},
  author={Rowan, Connar and others},
  year={2025},
  month=nov, pages={4576–4589} }

@article{Gr_bner_2020,
  title={Binary black hole mergers in AGN accretion discs: gravitational wave rate density estimates},
  volume={638},
  ISSN={1432-0746},
  url={http://dx.doi.org/10.1051/0004-6361/202037681},
  DOI={10.1051/0004-6361/202037681},
  journal={Astronomy \& Astrophysics},
  publisher={EDP Sciences},
  author={Gröbner, M. and others},
  year={2020},
  month=june, pages={A119} }

@article{LIGOScientific:2025snk,
    author = "Abac, A. G. and others",
    collaboration = "LIGO Scientific, Virgo, KAGRA",
    title = "{Open Data from LIGO, Virgo, and KAGRA through the First Part of the Fourth Observing Run}",
    eprint = "2508.18079",
    archivePrefix = "arXiv",
    primaryClass = "gr-qc",
    reportNumber = "LIGO-P2500167",
    doi = "10.3847/1538-4357/ae211e",
    journal = "Astrophys. J.",
    volume = "1004",
    number = "2",
    pages = "232",
    year = "2026"
}

@article{KAGRA:2023pio,
  author = "Abbott, R. and others",
  collaboration = "KAGRA, VIRGO, LIGO Scientific",
  title = "{Open Data from the Third Observing Run of LIGO, Virgo, KAGRA, and GEO}",
  eprint = "2302.03676",
  archivePrefix = "arXiv",
  primaryClass = "gr-qc",
  reportNumber = "LIGO-P2200316",
  doi = "10.3847/1538-4365/acdc9f",
  journal = "Astrophys. J. Suppl.",
  volume = "267",
  number = "2",
  pages = "29",
  year = "2023"
}

@article{LIGOScientific:2019lzm,
  author = "Abbott, Rich and others",
  collaboration = "LIGO Scientific, Virgo",
  title = "{Open data from the first and second observing runs of Advanced LIGO and Advanced Virgo}",
  eprint = "1912.11716",
  archivePrefix = "arXiv",
  primaryClass = "gr-qc",
  reportNumber = "LIGO-P1900206",
  doi = "10.1016/j.softx.2021.100658",
  journal = "SoftwareX",
  volume = "13",
  pages = "100658",
  year = "2021"
}

@article{Kass:1995loi,
  author = "Kass, Robert E. and Raftery, Adrian E.",
  title = "{Bayes Factors}",
  doi = "10.1080/01621459.1995.10476572",
  journal = "J. Am. Statist. Assoc.",
  volume = "90",
  number = "430",
  pages = "773--795",
  year = "1995"
}

@article{Haiman_2017,
  title={Electromagnetic chirp of a compact binary black hole: A phase template for the gravitational wave inspiral},
  volume={96},
  ISSN={2470-0029},
  url={http://dx.doi.org/10.1103/PhysRevD.96.023004},
  DOI={10.1103/physrevd.96.023004},
  number={2},
  journal={Physical Review D},
  publisher={American Physical Society (APS)},
  author={Haiman, Zoltán},
  year={2017},
  month=July }

@article{Secunda_2020,
  title={Orbital Migration of Interacting Stellar Mass Black Holes in Disks around Supermassive Black Holes. II. Spins and Incoming Objects},
  volume={903},
  ISSN={1538-4357},
  url={http://dx.doi.org/10.3847/1538-4357/abbc1d},
  DOI={10.3847/1538-4357/abbc1d},
  number={2},
  journal={The Astrophysical Journal},
  publisher={American Astronomical Society},
  author={Secunda, Amy and others},
  year={2020},
  month=Nov, pages={133} }

@article{Planck_2020,
  title={<i>Planck</i> 2018 results: VI. Cosmological parameters},
  volume={641},
  ISSN={1432-0746},
  url={http://dx.doi.org/10.1051/0004-6361/201833910},
  DOI={10.1051/0004-6361/201833910},
  journal={Astronomy \& Astrophysics},
  publisher={EDP Sciences},
  author={Aghanim, N. and others},
  year={2020},
  month=Sept, pages={A6} }

@ARTICLE{inayoshi_tamanini_2017,
       author = {{Inayoshi}, Kohei and {Tamanini}, Nicola and {Caprini}, Chiara and
         {Haiman}, Zolt{\'a}n},
        title = "{Probing stellar binary black hole formation in galactic nuclei via the imprint of their center of mass acceleration on their gravitational wave signal}",
      journal = {\prd},
         year = "2017",
        month = "Sep",
       volume = {96},
       number = {6},
          eid = {063014},
        pages = {063014},
          doi = {10.1103/PhysRevD.96.063014},
archivePrefix = {arXiv},
       eprint = {1702.06529},
 primaryClass = {astro-ph.HE},
       adsurl = {https://ui.adsabs.harvard.edu/abs/2017PhRvD..96f3014I}
}

@ARTICLE{lazarow_leslie_2024,
       author = {{Lazarow}, Malcolm and {Leslie}, Nathaniel and {Dai}, Liang},
        title = "{Gravitational waveform model for detecting accelerating inspiraling binaries}",
      journal = {\prd},
         year = 2024,
        month = oct,
       volume = {110},
       number = {8},
          eid = {083008},
        pages = {083008},
          doi = {10.1103/PhysRevD.110.083008},
archivePrefix = {arXiv},
       eprint = {2401.04175},
 primaryClass = {gr-qc},
       adsurl = {https://ui.adsabs.harvard.edu/abs/2024PhRvD.110h3008L}
}

@ARTICLE{tamanini_klein_2019,
       author = {{Tamanini}, Nicola and {Klein}, Antoine and {Bonvin}, Camille and
         {Barausse}, Enrico and {Caprini}, Chiara},
        title = "{Peculiar acceleration of stellar-origin black hole binaries: Measurement and biases with LISA}",
      journal = {\prd},
         year = 2020,
        month = mar,
       volume = {101},
       number = {6},
          eid = {063002},
        pages = {063002},
          doi = {10.1103/PhysRevD.101.063002},
archivePrefix = {arXiv},
       eprint = {1907.02018},
 primaryClass = {astro-ph.IM},
       adsurl = {https://ui.adsabs.harvard.edu/abs/2020PhRvD.101f3002T}
}

@ARTICLE{vijaykumar_tiwari_2023,
       author = {{Vijaykumar}, Aditya and {Tiwari}, Avinash and {Kapadia}, Shasvath J. and {Arun}, K.~G. and {Ajith}, Parameswaran},
        title = "{Waltzing Binaries: Probing the Line-of-sight Acceleration of Merging Compact Objects with Gravitational Waves}",
      journal = {\apj},
         year = 2023,
        month = sep,
       volume = {954},
       number = {1},
          eid = {105},
        pages = {105},
          doi = {10.3847/1538-4357/acd77d},
archivePrefix = {arXiv},
       eprint = {2302.09651},
 primaryClass = {astro-ph.HE},
       adsurl = {https://ui.adsabs.harvard.edu/abs/2023ApJ...954..105V}
}

@ARTICLE{yang_han_2025,
       author = {{Yang}, Shu-Cheng and {Han}, Wen-Biao and {Tagawa}, Hiromichi and {Li}, Song and {Zhang}, Chen},
        title = "{Indication for a Compact Object Next to a LIGO─Virgo Binary Black Hole Merger}",
      journal = {\apjl},
         year = 2025,
        month = aug,
       volume = {988},
       number = {2},
          eid = {L41},
        pages = {L41},
          doi = {10.3847/2041-8213/adeaad},
archivePrefix = {arXiv},
       eprint = {2401.01743},
 primaryClass = {astro-ph.HE},
       adsurl = {https://ui.adsabs.harvard.edu/abs/2025ApJ...988L..41Y}
}

@ARTICLE{zhao_yan_2026,
       author = {{Zhao}, Xinmiao and {Yan}, Han and {Chen}, Xian},
        title = "{A Novel Method to Construct Frequency-Domain Gravitational Waveform for Accelerating Sources}",
      journal = {arXiv e-prints},
         year = 2026,
        month = mar,
          eid = {arXiv:2604.00253},
        pages = {arXiv:2604.00253},
          doi = {10.48550/arXiv.2604.00253},
archivePrefix = {arXiv},
       eprint = {2604.00253},
 primaryClass = {astro-ph.HE},
       adsurl = {https://ui.adsabs.harvard.edu/abs/2026arXiv260400253Z}
}

\end{document}